\documentclass{aa}
\usepackage{epsfig}
\usepackage{amssymb}
\begin{document}

  \thesaurus{06 (08.06.3; 08.15.1; 08.22.3; 03.20.4)}

  \title{A search for new $\gamma$ Doradus stars in the Geneva
                photometric database}

  \author{L. Eyer
         \and
         C. Aerts \thanks{Postdoctoral Fellow,
                 Fund for Scientific Research, Flanders}}

  \institute{Instituut voor Sterrenkunde,
             Katholieke Universiteit Leuven,
             Celestijnenlaan 200 B, B-3001 Leuven,
             Belgi\"e}

  \offprints{L.Eyer}
  \mail{Laurent.Eyer@ster.kuleuven.ac.be}

  \date{Received / Accepted}

  \titlerunning{A search for new $\gamma$ Dor stars in the Geneva
                photometric database}
  \authorrunning{L. Eyer \and C. Aerts}
  \maketitle

\begin{abstract}
We present a study of selected potential $\gamma\,$Doradus stars observed in
the Geneva photometric system. Eleven stars were monitored at least during
the last year of functioning of the 70 cm Swiss telescope at La Silla, leading
to about one thousand photometric measurements and permitting a detailed
analysis of these stars. After the analysis, five stars are thought to be
constant stars. We report three new $\gamma$ Dor stars. A spectroscopic campaign
has been started at La Silla with the new 1m20 Swiss telescope, equipped with
the spectrograph CORALIE, of which we show the results for the selected stars.
\end{abstract}

\section{Introduction}
The $\gamma\,$Dor stars constitute a new class of pulsating variables that was
discovered only recently. Krisciunas (\cite{Krisciunas}) and Zerbi
(\cite{Zerbi}) review the history of the discovery and the current
observational status of this group of variables. We refer to these papers for a
general introduction on $\gamma\,$Dor stars, but we mention here that these
objects are thought to exhibit multiperiodic high-order g-mode pulsations with
periods in the range of 0.5 - 3 days.

Much effort is currently made to find new members of this group, in order
to constrain their pulsational characteristics and their position in the
HR\,diagram. This is especially relevant since the subject of the precise
excitation mechanism is not yet settled. Promising theoretical models are
currently being worked out. No direct confrontation between the new theoretical
ideas about the driving and the observational status has been performed yet, 
one of the reason being that the number of {\it bona fide} members is still low
and that these stars have quite a large variety in their observational
behaviour.

We already performed a systematic detection and classification of $\gamma\,$Dor
stars from the Hipparcos Periodic Annex using the Geneva photometric system.
The method used was a multivariate discriminant analysis and it led to the
discovery of 14 new $\gamma\,$Dor candidates (Aerts et al. \cite{Aerts&al}).
Moreover, Handler (\cite{Handler}), searching for new $\gamma\,$Dor stars in the
Hipparcos catalogues of the periodic and unsolved variables, found respectively
13 and  18 new ``prime'' $\gamma\,$Dor candidates which are non-redundant
with other studies.

The paper is organised as follows. In Sect.\,2, we present the obtained data
from the Geneva photometry, and compare it to the Hipparcos photometry. In
Sect.\,3, we search for periodic behaviour in the data. In Sect.\,4, we
present the spectra of the stars. Finally, we end with some conclusions in
Sect.\,5.

\section{Geneva Photometric data}
The content of the Geneva catalogue is the union of more than 200
scientific programmes, including all stars in the Bright Star Catalogue south of
$\delta = +20^\circ$. The Geneva photometric database now contains 344~500
measurements of 47~600 stars. There are seven filters, namely U, B, V, B1, B2,
V1, G. During the reduction scheme, a weight is assigned to each measurement
and the noise level of single measurements with the weights 3 or 4 for the V
band is about 4 to 5 milli-magnitudes. For the article, we take the
intermediate value of 4.5 milli-magnitudes.

We started a search in the Geneva photometric database for
$\gamma\,$Dor candidates. 
We selected in the Geneva catalogue F0-F9 stars which fitted the
observational window and with a high dispersion, $\sigma_{m_V}$. The eleven
chosen candidates were monitored with the P7 photometer attached to the 70cm
Swiss telescope at La Silla observatory. We here report our analyses of
the new photometric data.

Generally, we disregarded the very few measurements that already existed of the
programmed stars in the Geneva database and used only the data taken during
the dedicated campaigns. These consist from 2 to 7 runs of each 3 weeks,
during which each star was observed typically twice per night. The total time
span of the data varies between 1 to 5 years with a mean of 2 years.

The Geneva photometry permits an accurate determination of temperature
$T_{\mbox{\tiny eff}}$, gravity $\log(g)$, and metallicity [M/H]. In
Table~\ref{tab:anphys} we list the physical properties of our 11 stars derived
from the K\"unzli et al. (\cite{Kunzli}) calibrations. The errors given are the
ones resulting from the interpolation in the tables of standard stars. They are
only valid assuming that the average colours are free from errors. The standard
errors should therefore not be considered as physical error estimates. Our
targets are in general slightly metal weak and some are just slightly hotter
than the sun.

\begin{table*}
 \caption{\label{tab:anphys} Basic statistical photometric information and 
          basic physical properties derived from the Geneva photometry}
 \begin{center}
   \begin{tabular}{|c|c|c|c|c|c|c|c|c|} \hline
HD     &Sp. Type &$n_G$&$\overline{m_V}$&$\sigma_{\overline{m_V}}$&$p_{\mbox{\tiny val}}$&$T_{\mbox{\tiny eff}}$&$\log(g)$&[M/H]\\ \hline
  5590 & F2V     & 102 & 9.1918 & .0004 & .958    & 6640 $\pm$ 40 & 4.32 $\pm$ 0.14 & -0.11 $\pm$ 0.09\\
  7455 & F5V     &  43 & 8.4212 & .0006 & .779    & 6380 $\pm$ 40 & 3.93 $\pm$ 0.13 & -0.15 $\pm$ 0.09\\
  8393 & F7w...  & 104 & 9.4761 & .0004 & .548    & 5980 $\pm$ 30 & 3.89 $\pm$ 0.13 & -0.40 $\pm$ 0.10\\
 10167 & F0V     & 119 & 6.6575 & .0008 & .000    & 6950 $\pm$ 60 & 4.44 $\pm$ 0.12 & -0.29 $\pm$ 0.10\\
 12901 & F0      & 122 & 6.7238 & .0015 & .000    & 7010 $\pm$ 60 & 4.47 $\pm$ 0.09 & -0.40 $\pm$ 0.11\\
 22001 & F5IV-V  &  34 & 4.7008 & .0007 & .789    & 6640 $\pm$ 40 & 4.34 $\pm$ 0.14 & -0.19 $\pm$ 0.09\\
 26298 & F0/F2V  & 107 & 8.1451 & .0005 & .000    & 6720 $\pm$ 50 & 4.36 $\pm$ 0.16 & -0.33 $\pm$ 0.10\\
 27604 & A8V+... & 111 & 6.0663 & .0004 & .629    & 6310 $\pm$ 30 & 3.60 $\pm$ 0.13 &  0.02 $\pm$ 0.08\\
 33262 & F7V     & 94  & 4.6974 & .0007 & .000    & 6160 $\pm$ 30 & 4.58 $\pm$ 0.13 & -0.22 $\pm$ 0.09\\
 35416 & F3V     &  96 & 7.5216 & .0006 & .001    & 6600 $\pm$ 40 & 4.39 $\pm$ 0.14 & -0.46 $\pm$ 0.11\\
 48501 & F2V     & 140 & 6.2381 & .0014 & .000    & 7020 $\pm$ 60 & 4.49 $\pm$ 0.10 & -0.10 $\pm$ 0.10\\ \hline  
  \end{tabular}
 \end{center}
\end{table*}

\subsection{Variability detection}
A $\chi^2$ test was used to evaluate the variability status of the photometric
time series. Table~\ref{tab:anphys} lists also the main characteristics of
these photometric time series. Because there is usually a huge time gap between
the first data ever taken in the Geneva system and the newer dedicated
photometric sequences, the measurements taken before the Julian date
$2\,448\,000$ were removed.
All more recent data points were kept, such that we have a higher resolution
in the frequency domain as will become evident from the subsequent period
searches. In Table~\ref{tab:anphys}, $n_G$ is the number of measurements with
weight strictly greater than $2$, $\overline{m_V}$ is the mean $V$ magnitude,
$\sigma_{\overline{m_V}}$ is the dispersion of the mean $V$ magnitude,
$p_{\mbox{val}}$ is the p-value associated with the $\chi^2$.

In the sample of the 11 stars studied, fixing the size of the test at 1\%, we
found that five stars are constant stars (HD\,5590, HD\,7455, HD\,8393,
HD\,22001, HD\,27604). In these stars the high initial $\sigma_{m_V}$ was due
to  off measurements at the beginning of the time series.  We note that the
time series of the star HD\,8393 seems to show a slight drift and that the time
series of the star HD\,27604 has a few data points with a high magnitude.

\subsection{Cross verification with Hipparcos data}
\begin{figure}[htbp]
  \centering%
  \mbox{\psfig{file=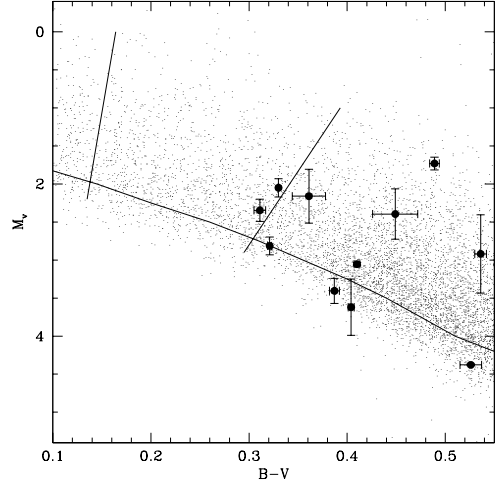,height=70mm,width=70mm}}%
  \caption{\label{fig:hrdiag} HR diagram for the eleven stars based on the
           Hipparcos parallax and the $B-V$ index from the Hipparcos catalogue.
           The solid lines indicate the main sequence and the borders of the
           instability strip. The small points are the Hipparcos stars with
           very accurate parallaxes and $B-V$ colours.}
\end{figure}

The 11 stars were also measured by the Hipparcos satellite. In
Table~\ref{tab:anhip} we have the correspondence between HD and HIP (Hipparcos
identifier), the $V$ magnitude, the absolute magnitude $M_V$, its error
$\sigma_{M_V}$, the colour $B-V$ and its associated error $\sigma_{B-V}$,
the number of observations $n_H$, the median magnitude $H\!p$, the standard
error of the $H\!p$ median magnitude $\sigma_{H\!p}$, the p-value associated
with the $\chi^2$ test $p_{\mbox{val}}$ and the result of the analysis by the
Hipparcos teams RH, which indicates that the star was classified as:
constant C, microvariable M, double star D, member of the atlas of the unsolved
variable U (cf. ESA \cite{ESA} and Eyer \cite{Eyer}). The HR diagram is
represented in Fig.~\ref{fig:hrdiag} with the location of the eleven stars.
Note that there is a probable identification problem in the Hipparcos catalogue.
The star HD\,7455 is referred to HIP\,5745. According to its absolute magnitude
and to its TYCHO colours, we conclude that the correct identification of this
star is HIP\,5735 and not HIP\,5745.

\begin{table*}
 \caption{\label{tab:anhip} Magnitude V, absolute $M_V$ magnitude
          deduced from Hipparcos parallax and result of the
          Hipparcos main mission photometric analysis}
 \begin{center}
  \begin{tabular}{|c|c|c|c|c|c|c|c|c|c|c|c|} \hline
   HIP&      HD& V    &$M_V$ & $\sigma_{M_V}$&$B-V$&$\sigma_{B-V}$&$n_H$& $H\!p$&$\sigma_{H\!p}$& $p_{\mbox{\tiny val}}$& RH \\ \hline
   4481&   5590& 9.21 & 3.62 & 0.37 & 0.404 &  0.003& 116 & 9.2992 & .0013&0.420 &   \\
   5735&   7455& 8.44 & 2.39 & 0.33 & 0.449 &  0.023& 131 & 8.5365 & .0017&0.096 & C \\
   6387&   8393& 9.49 & 2.92 & 0.51 & 0.536 &  0.006& 122 & 9.6049 & .0020&0.001 &   \\
   7649&  10167& 6.67 & 2.05 & 0.12 & 0.330 &  0.000& 174 & 6.7527 & .0008&0.000 &   \\
   9807&  12901& 6.74 & 2.35 & 0.15 & 0.311 &  0.006& 134 & 6.8180 & .0014&0.000 & U \\
   16245& 22001& 4.71 & 3.05 & 0.02 & 0.410 &  0.003& 130 & 4.8043 & .0004&0.946 & C \\
   19383& 26298& 8.16 & 2.16 & 0.35 & 0.361 &  0.017&  86 & 8.2464 & .0015&0.678 & C \\
   20109& 27604& 6.08 & 1.73 & 0.08 & 0.489 &  0.005& 117 & 6.1838 & .0005&0.834 & D \\
   23693& 33262& 4.71 & 4.38 & 0.01 & 0.526 &  0.011& 100 & 4.8191 & .0008&0.000 &   \\
   25183& 35416& 7.53 & 3.40 & 0.17 & 0.387 &  0.005& 133 & 7.6039 & .0015&0.000 & M \\
   32144& 48501& 6.26 & 2.81 & 0.12 & 0.321 &  0.002& 283 & 6.3324 & .0066&  --  & D \\ \hline
  \end{tabular}
 \end{center}
\end{table*}

We remark that out of the eleven stars, one has only flagged measurements
(double star), five stars can be declared to be constant, five can be declared
variable.  The classification from Hipparcos is nearly consistent with the
classification obtained from the Geneva photometry. However, there are two
discrepant cases: HD\,8393 (HIP\,6387) and HD\,26298 (HIP\,19383). The latter
case is explained by the fact that the precision of Hipparcos is not as good as
the Geneva photometry at these magnitudes; the former case is more puzzling.

\section{Period search}
For the 6 remaining stars left from the analysis of the Geneva photometry, we
searched for periodic behaviour with different methods, namely Fourier (Deeming
\cite{Deeming} and Ferraz-Mello \cite{FerrazMello}, Scargle \cite{Scargle}) and
PDM (Jurkevich \cite{Jurkevich}, Stellingwerf \cite{Stellingwerf}). The results
are listed in Table~\ref{tab:rechper}. The frequency resolution is taken as
$1/(8(t_{n_G}-t_1))$, where $t_1$ and $t_{n_G}$ are the times for the
first and last data points. The interval in frequency is chosen between 0 and
30 [1/days]. The reason for searching short periods is a possible contamination
of the sample by $\delta$ Scuti stars.

\begin{table}
 \caption{\label{tab:rechper} Results using different period search algorithms
          for the V band. The frequencies are given in cycle/day}
 \begin{center}
   \begin{tabular}{|c|c|c|c|c|} \hline
   HD     & Deeming  & Ferraz-Mello & Scargle  & PDM    \\ \hline
   10167  & 0.48760  & 1.00256     & 0.51258  & 0.51258\\
   12901  & 1.21552  & 1.21552     & 1.21551  & 1.21551\\
   26298  & 2.00061  & 1.00014     & 1.00199  & 2.00179\\
   33262  & 0.00952  & 0.00281     & 0.00952  & 0.00967\\
   35416  & 0.00467  & 0.00016     & 1.00464  & 1.97870\\
   48501  & 0.09122  & 0.09122     & 0.09126  & 0.09110\\ \hline
  \end{tabular}
 \end{center}
\end{table}

For two stars (HD\,12901, HD\,48501) there is an agreement between all methods.
For two other stars (HD\,10167, HD\,26298), the resulting frequencies are close
to  a sampling frequency. For the star HD\,33262, the period is about 100 days.
For the star HD\,35416, each method gives a different result. It seems
that the time series of this star shows a long term trend explaining the long
periods found. With a model with trend and noise the behaviour of the star is
explained (within a 1\% test size).

\begin{figure}[htbp]
  \centering%
  \mbox{\psfig{file=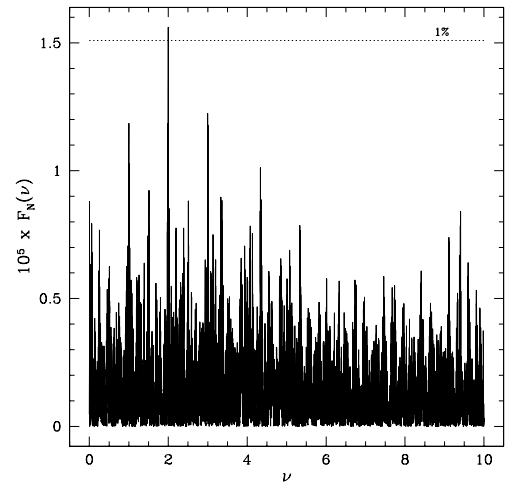,height=70mm,width=70mm}}%
  \caption{\label{fig:montecarlo} Power spectrum of HD\,26298,
            the horizontal line is the 1\% false alarm probability obtained
            from Monte-Carlo simulations.}
\end{figure}

For the time series, we performed extensive Monte-Carlo simulations to determine
the false alarm probability. The $H_0$ hypothesis is constructed as follows.
We take a gaussian signal sampled at the same times with the same variance,
and perform a Fourier transform over a frequency interval from 0 to 10
[1/days] with a time step of $1/(8(t_{n_G}-t_1))$. In the delicate case of the
time series of HD\,26298, we see that the maximum of the power
spectrum is higher than the 1\% level derived from simulations
(cf. Fig.~\ref{fig:montecarlo}),
which clarifies that this star is variable.

\begin{figure}[htbp]
  \centering%
  \mbox{\psfig{file=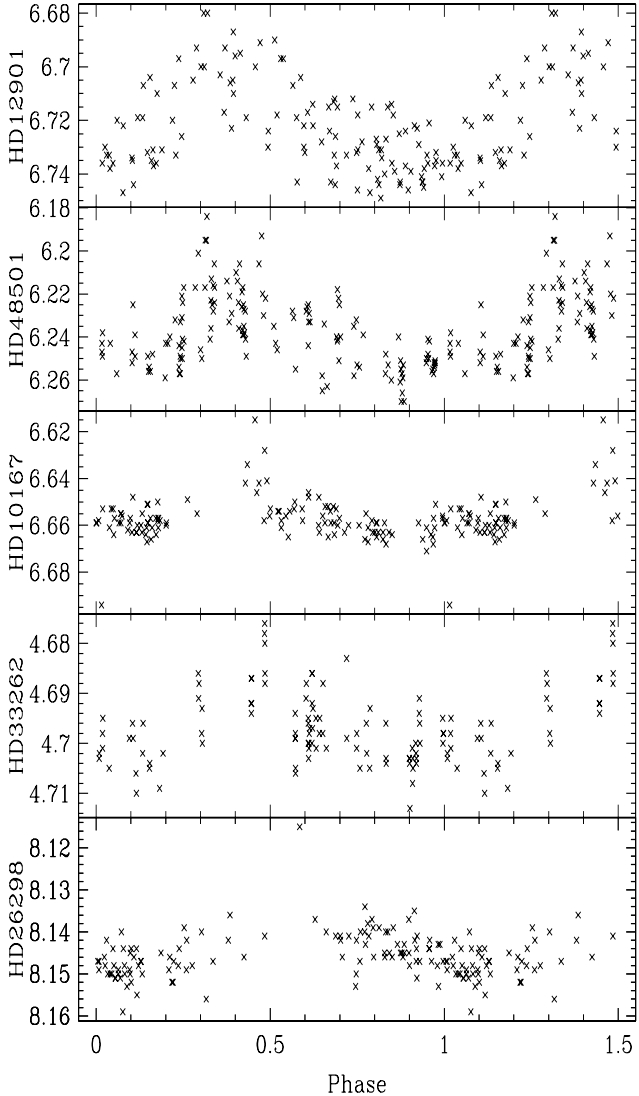,height=160mm,width=88mm}}%
  \caption{\label{fig:desph} Phase diagrams of the V magnitude for the stars
           HD\,12901, HD\,48501, HD\,10167, HD\,33262, HD\,26298, with periods
           taken from the results of the PDM method given in
           Table~\ref{tab:rechper}.}
\end{figure}

We present in Fig.~\ref{fig:desph} the folded curves with the period obtained
from the PDM method.

For the 6 stars for which a period search was performed, three stars are in the
compatible range for $\gamma$ Dor stars behaviour. However, two stars have
periods related to the sampling period of one day, leaving gaps in the phase
diagram (cf.~Fig.~\ref{fig:desph}).

The stars with a sufficiently dense covering over all phases have clearly high
residuals after prewhitening with the main period, pointing to possible
multiperiodic behaviour or a non-regular behaviour.
For the star HD\,12901, the fraction of the variance explained by a sinusoidal
model with one period (0.8227 day, amplitude: 16.3 milli-mag) is 49\%,
and with two periods (0.8227 and 0.8430 day, amplitudes: 15.1 and 9.7
milli-mag) is 67\%.
For the star HD\,48501, the fraction of the variance explained by a sinusoidal
model with one period (10.959 days, amplitude: 15.7 milli-mag) is 44\%,
and with two periods (10.959 and 0.7750 day, amplitudes: 14.4 and 12.8
milli-mag) is 70\%. Since there is a period close to 1 day, we can include
this star in our list of candidates. We conclude that both HD\,12901 and
HD\,48501 are clearly multiperiodic.

\subsection{Pulsation, eclipses ?}
The sample might be contaminated by EW eclipsing binaries, the period of which
can be of the same order than that of the intrinsic periods of  $\gamma\,$Dor
stars.

A test that discriminates between these possibilities (besides spectroscopic
observations) is a study of the periodic signal in the colour. The bona-fide
$\gamma\,$Dor stars are known to have colour changes. These changes have the
same periods than those found in the light variability.

If the system is eclipsing and if the two stars are identical, then no colour
variations will be observed. If, however, the two stars have different colours,
then the signal will have a double periodic behaviour.

For HD\,12901, we find the same period in the colour $B-V$ as in $V$.
For HD\,48501, the period found is not near 10 days but is 0.91 day. This points
out that the period around 10 days might be an alias period of the one near one
day. For the stars HD\,10167 and HD\,26298, it seems that the period is half of
the period found in the $V$ magnitude. HD\,35416 has no significant colour
variations.

\subsection{Amplitudes in different filters}
We performed the period search in different filters (U, B, B1)
to check if the main period mentioned above appears
in all these filters. 
The frequency spectra in the different filters behave as follows:
\begin{itemize}
 \item for HD\,12901 and HD\,48501, the spectra in the different bands are
       similar. The highest amplitude for both occurs in the B1 filter.
 \item for HD\,10167, there is a different spectrum in the U band but all the
       other bands have the same structure (B1 having the highest amplitude).
 \item HD\,33262, HD\,35416 have similarities in certain different
       bands.
 \item HD\,26298 has a different behaviour in different bands.
\end{itemize}

\subsection{Conclusion on the photometric variability analysis}
Out of the eleven stars five are constant, one has no significant
periodic behaviour (HD\,35416), one has a long term behaviour (HD\,33262),
one could be a $\gamma\,$Dor star but there is an aliasing problem (HD\,48501),
and three stars have periods compatible with $\gamma\,$Dor behaviour. Two of
these have an incomplete phase coverage (HD\,10167, HD\,26298) and one has a
good phase coverage (HD\,12901). The check for colour variations does not
permit to rule out any cases.

\section{CORALIE measurements}
\subsection{Data description}
\begin{figure}[htbp]
  \centering%
  \mbox{\psfig{file=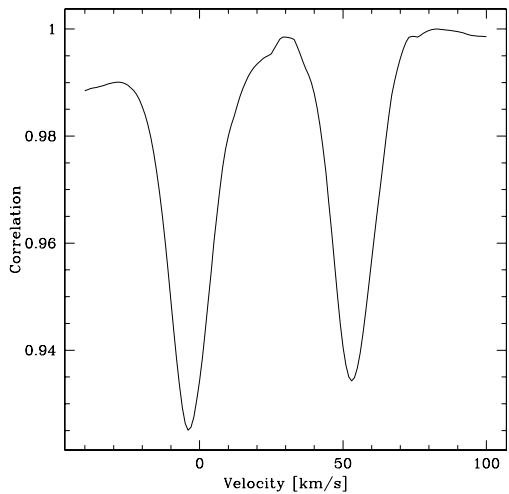,height=70mm,width=70mm}}%
  \caption{\label{fig:cordble} Correlation for the star HD\,10167. It is
             clearly a double star.}
\end{figure}

We have started a campaign with the CORALIE spectrograph for detecting
possible line-profile variations in the $\gamma\,$Dor candidates. CORALIE is an
echelle spectrograph attached to the new 1m20 Swiss telescope at La Silla, which
was primarily built to search for extraterrestrial planets. The software gives
online radial velocities of the observed object by doing a cross
correlation with a template. The correlation profile is directly accessible,
and so can be used to derive immediately possible duplicity and to determine
the rotational velocity of the star.
The correlation profile reflects also the double star character by
presenting a double peak. The star HD\,10167 is clearly a double star
(cf. Fig.~\ref{fig:cordble}). The photometric variability of this object is
probably due to the duplicity and not to a $\gamma$\,Dor phenomenon. We find
that the star HD\,27604 is also a double star.

Besides the correlation profile, the spectrum itself can be obtained from
388\,nm to 681\,nm with a resolution of about 50\,000 at 500\,nm. We present
here the spectra of nine stars. Two stars were too faint to be measured.


We searched for a spectral line which is not blended and which could show clear
profile variations. For each star a set of five spectra was taken and for the
promising candidates larger sets of spectra have been taken and will be
presented in the near future. The question of chromospheric activity will also
be addressed. We will elaborate much more extensively on our line-profile
variation study of many candidate $\gamma$\,Dor stars in a forthcoming paper.

In order to select a suitable spectral line, we use the spectrum of a standard
star, with a high resolution. In this way, we are then able to distinguish
which lines are not blended. Our choice was to concentrate on the calcium line
at 6122 \AA.  Other spectral lines can be taken as the FeII at 4508 \AA ~and
also TiII at 4501 \AA. Many stars show a rapid rotation which broadens the
lines and makes an analysis difficult.

\begin{figure}[htbp]
  \centering%
  \mbox{\psfig{file=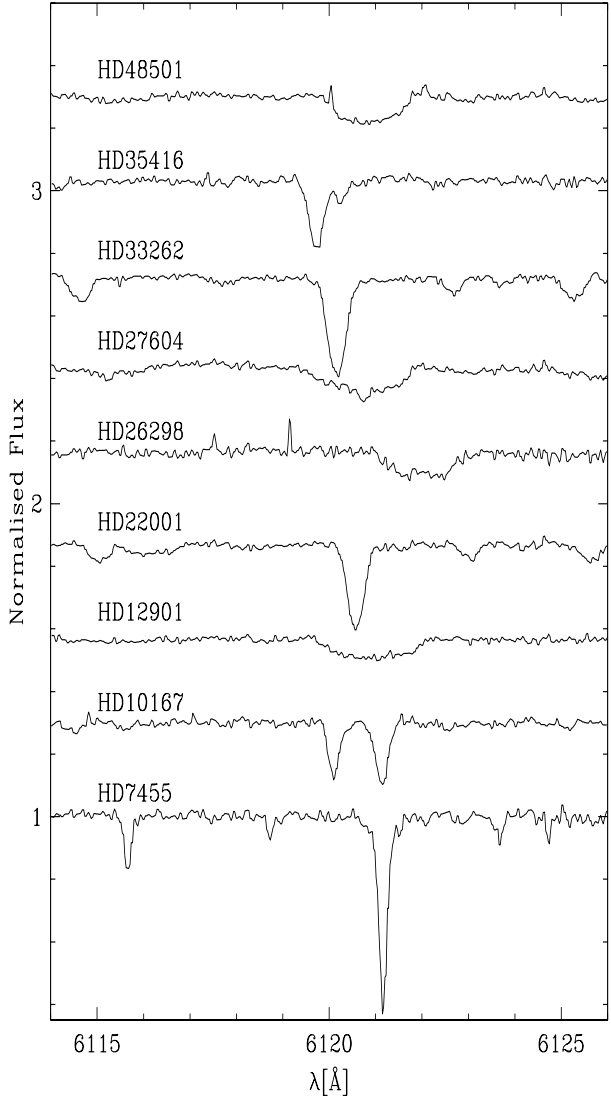,height=160mm,width=88mm}}%
  \caption{\label{fig:raieca} CORALIE spectra for the 9 stars observed
   around the unblended CaI line at 6122 \AA. Two stars in our list are
   too faint to be observed.}
\end{figure}

We superposed the spectra of the 9 stars around the unblended Ca line in
Fig.~\ref{fig:raieca}.

\subsection{Rotation speed and rotation period}
We determined the $v\sin(i)$ by the following method: we used the star HD\,7455
which has a small $v\sin(i)$ as a template and applied an artificial rotation
broadening to its spectrum with different $v\sin(i)$. We computed then its
correlation profile. We compared the correlation profile of the star for which
we want to know the $v\sin(i)$ with the different computed correlation
templates. The solution which minimised the residuals between the solution and
the template is taken as the best $v\sin(i)$. They are listed in
Table~\ref{tab:rot}. The underlying hypothesis is that all the broadening is
due to rotation. The radii and the masses were determined by doing an interpolation
using the tables of Schaller et al. (\cite{Schaller}), and Schaerer et al.
(\cite{Schaerer}) and by taking the physical parameters from the Geneva
photometry. We then determined the maximal rotation periods of the stars by
assuming $i=\pi/2$. We note that the estimates of the rotation period are of
the same order of magnitude as the periods found in the photometry. It is
therefore important to find multiperiodicity and/or additional evidence of the
$\gamma\,$Dor nature to be able to exclude rotational modulation as a cause of
the observed variability.

\begin{table}
 \caption{\label{tab:rot} Rotation velocities $v\sin(i)$, radii, masses and
                          rotation periods of the sample}
 \begin{center}
   \begin{tabular}{|c|c|c|c|c|} \hline
   HD     & vsini [km/s] & $R/R_{\odot}$&$M/M_{\odot}$& Period [days]\\ \hline
   12901  & 64           &   1.2        & 1.5 &  0.9\\
   22001  & 10           &   1.3        & 1.4 &  6.4\\
   26298  & 50           &   1.2        & 1.4 &  1.2\\
   33262  & 8            &   0.9        & 1.2 &  5.7\\
   35416  & $<$ 4        &   1.1        & 1.3 &  -- \\
   48501  & 40           &   1.1        & 1.5 &  1.4\\  \hline
  \end{tabular}
 \end{center}
\end{table}

\section{Conclusion}
The more thoroughly the candidates were studied the more they were rejected.
We end therefore with a shorter list of $\gamma\,$Dor stars. Establishing
such a list needs strong efforts, but it is important for describing precisely
the physical properties of this pulsating star group.

We have found three new $\gamma\,$Doradus stars (HD\,12901, HD\,48501,
HD\,26298). One of them (HD\,12901) was indicated by Handler
(\cite{Handler}) to have a possible $\delta\,$Scuti behaviour on the basis of
Hipparcos data. Our Geneva photometry, however, clearly points towards multiple
periods longer than those of $\delta\,$Scuti stars and in the region expected
for $\gamma\,$Doradus stars. HD\,48501 is also a multiperiodic variable. The
line-profile variability of these three new $\gamma\,$Dor stars will be
discussed in a separate paper.

It is presently important to find as many confirmed $\gamma\,$Doradus stars as
possible, in order to constrain the suggested theoretical excitation models. In
particular, the question remains whether or not the $\gamma\,$Doradus
phenomenon only occurs for F\,0 -- F\,2 type stars or also for much cooler
objects as suggested by Aerts et al.\ (\cite{Aerts&al}). If the latter is true,
then it is tempting to assume that the $\gamma\,$Doradus stars may exhibit
simultaneously opacity-driven g-modes and stochastically excited solar-like
p-modes. The occurrence of both such types of pulsations would broaden the
interest for the stars because they would become unique objects for seismology.
In this respect, the search for solar-like oscillations in bright selected
$\gamma\,$Doradus stars by future seismology space missions, such as MOST,
MONS and COROT, would be extremely valuable.

\acknowledgements{We would like to thank very warmly Claudio Melo for his very
kind and efficient help.}

\end{document}